%% file: TopDileptonAfb_PRL.tex
\documentclass[aps,prl,reprint,amsmath,amssymb,showpacs,superscriptaddress,floatfix]{revtex4-1}
 \def\LineNumbers{false}
 \def\AuthorList{true}

\usepackage{ifthen}
\usepackage{lineno}
\usepackage{amsmath}    
\usepackage{subfigure} 
\usepackage{amssymb}
\usepackage{epsfig}
\usepackage{ifthen}
\usepackage{rotating}
\usepackage{ctable}
\usepackage{graphicx}
\usepackage{subfigure}
\usepackage{dcolumn}
\usepackage{bm}
\usepackage{footnote}
\usepackage{multirow}
\usepackage{textcomp}
\usepackage{mathrsfs}
\usepackage[utf8]{inputenc}

\usepackage{color}
\usepackage{eso-pic} 
 \DeclareGraphicsExtensions{.eps,.ps,.eps.gz,.ps.gz}    
 
\usepackage{xspace}
\usepackage{fancyhdr}

\makeatletter

\renewcommand*{\p@subsection}{}

\renewcommand*{\p@subsubsection}{}
\makeatother

\input{afbcommands}
\makeatletter\def\ps@titlepage{\ps@fancy}\makeatother

\ifthenelse{\equal{\AuthorList}{true}}{
        \rhead{}
        \lhead{}
}{ 
        \rhead{CDF/ANAL/TOP/CDFR/11052 \\ version 3.0}
        \lhead{\includegraphics[height=2cm]{img/cdfii}}
}

\ifthenelse{\equal{\LineNumbers}{true}}{
        \linenumbers

        \def\equation{\begin{linenomath*}\begin{oldeqn}}
        \def\endequation{\end{oldeqn}\end{linenomath*}\hspace*{-0.15cm}}
}{}
\ifthenelse{\equal{\LineNumbers}{true}}{
        \linenumbers

        \def\subequations{\begin{linenomath*}\begin{oldsubeqns}}
        \def\endsubequations{\end{oldsubeqns}\end{linenomath*}\hspace*{-0.3cm}}
}{}

\begin{document}
\input{tables/PublicTables.tex}

\title{Measurement of the inclusive leptonic asymmetry in top-quark pairs that decay to two charged leptons at CDF}

\ifthenelse{\equal{\AuthorList}{true}}{
        \input{author}
}{ 
        \vspace*{2cm}
        \author{The CDF Collaboration} \affiliation{URL http://www-cdf.fnal.gov}
}

\date{\today}

\begin{abstract}
\ifthenelse{\equal{\AuthorList}{false}}{
\vspace*{0.2cm}
}{}

We measure the inclusive forward-backward asymmetry of the charged-lepton pseudorapidities from top-quark pairs produced in proton-antiproton collisions, and decaying to final states that contain two charged leptons (electrons or muons), using data collected with the Collider Detector at Fermilab. With an integrated luminosity of 9.1~$\fbm$, the leptonic forward-backward asymmetry, $\afblep$, is measured to be $0.072 \pm 0.060$ and the leptonic pair forward-backward asymmetry, $\afbdeta$, is measured to be $0.076 \pm 0.082$, compared with the standard model predictions of $\afblep = 0.038 \pm 0.003$ and $\afbdeta = 0.048 \pm 0.004$, respectively. Additionally, we combine the $\afblep$ result with a previous determination from a final state with a single lepton and hadronic jets and obtain $\afblep = 0.090^{+0.028}_{-0.026}$.
\end{abstract}

\maketitle

\setlength{\headheight}{0pt}


Recent measurements of the forward-backward asymmetry ($\afb$) of the rapidity difference of top anti-top ($\ttbar$) quark pairs ($\afbtt$) production in proton-antiproton collisions with center-of-mass energy of 1.96~TeV at the Fermilab Tevatron collider~\cite{Aaltonen:2012it,d0_afb_prd2011,CDF:2013gna} show deviations from the predictions from the standard model (SM) of particle physics~\cite{Bernreuther:2012sx}. This is of significant interest as the SM predicts only small asymmetry due to interference among diagrams starting at next-to-leading order (NLO), while non-SM particles or interactions could modify the $\afbtt$ in a larger range~\cite{Jung:2010yn,*Jung2010238,*Frampton2010294,*jhep05(2011)070,*Chen:2010hm,*PhysRevD.82.094011,*PhysRevD.82.071702,
*PhysRevD.82.094009,*PhysRevD.82.034026,*PhysRevD.81.114004,*PhysRevD.81.055009,*PhysRevD.81.015004,
*PhysRevD.81.034012,*PhysRevD.82.034034,*PhysRevD.81.014016,*PhysRevD.81.113009,*PhysRevD.78.094018,
*PhysRevD.80.051701,*jhep11(2010)039,*Cheung2009287}. A separate set of useful observables relies on the pseudorapidities of the charged leptons that can originate from the cascade decays of the top quarks. These include the $\afb$ in the charge-weighted pseudorapidities of the charged lepton($\ell$, where we only consider electrons and muons), the so-called leptonic $\afb$ ($\afblep$), and the leptonic pair $\afb$ ($\afbdeta$) for the final state with two charged leptons (dilepton final state), defined with the pseudorapidity difference between the two charged leptons~\cite{Bernreuther201090}. For example, the resonant production of $\ttbar$ pairs via a hypothetical gluon with axial couplings (``axigluon") could cause the $\afbtt$ to deviate from its SM value; various axigluon couplings to the top quarks could produce the same value of $\afbtt$, but with very different values of $\afblep$ and $\afbdeta$~\cite{PhysRevD.87.034039}. Measurements of $\afb$ of the leptons also have the experimental advantage of exploiting the precisely measured angles of the lepton trajectories, which reduces systematic uncertainties on the final observables~\cite{Aaltonen:2013vaf}.

In this Letter, we summarize the measurements of the $\afblep$ and the $\afbdeta$ in the dilepton final state using the data collected by the CDF II detector, corresponding to the full Tevatron Run II data set, which corresponds to an integrated luminosity of 9.1~$\fbm$~\cite{ZHThesis}. Additionally, we report on the most sensitive measurement of $\afblep$ from the CDF collaboration by combining the new measurement of the $\afblep$ with the previous measurement~\cite{Aaltonen:2013vaf} in the final state involving one lepton and jets (lepton+jets final state). All results are {\it inclusive} in that they are extrapolated to the full pseudorapidity range.

The CDF II detector, described in detail in Ref.~\cite{PhysRevD.71.032001,geometry}, is a general-purpose particle detector employing a large charged-particle tracking volume inside a solenoidal magnetic field coaxial with the beam direction, surrounded by calorimeters and muon detectors. A sample enriched in $\ttbar$ events yielding dilepton final states ($\ttbar\rightarrow \ell^{+}\ell^{-} \nu  \bar{\nu} b \bar{b}$) is selected by requiring two oppositely charged leptons with $\pt > 20~\gevc$, two narrow clusters of energy deposit in the calorimeters, corresponding to collimated clusters of incident hadrons (jets), and an imbalance in the total event transverse momentum (missing transverse energy~\cite{MET}, or $\met$) that is consistent with the presence of two neutrinos. Specifically, we require events to pass the same requirements that were used in the measurement of the $\ttbar$ cross section ~\cite{Aaltonen:2013tsa}, except for the additional requirement that at least one jet have the signature of originating from $b$-quark fragmentation. We also raise the minimum dilepton invariant mass requirement from 5 to 10~$\gevcc$ to reduce background modeling uncertainties.

Several physical processes mimic the signature of top-quark pairs in the dilepton final state, such as production of $Z$ boson or a virtual photon in association with jets ($Z/\gamma^{*}$+jets), production of $W$ boson with jets ($W$+jets), diboson production ({\it WW}, {\it WZ}, {\it ZZ} and $W\gamma$), and $\ttbar$ production where one of the $W$ bosons from the top-quark pair decays hadronically and one jet is misidentified as a lepton. The estimation of background and SM $\ttbar$ signal is based on the same method of Ref.~\cite{Aaltonen:2013tsa}, which exploits both Monte Carlo (MC) simulations and data-based techniques. For the MC simulations, leading-order event generators are configured to use the CTEQ6.1L set of parton-distribution functions (PDFs), while NLO event generators use CTEQ6.1M. \textsc{Pythia}~\cite{Sjostrand:2006za} is used for modeling the parton hadronization; a \textsc{geant}-based simulation~\cite{Gerchtein:2003ba,Brun:1987ma} is used to model the detector response. A $\ttbar$ sample generated using the \textsc{powheg} generator~\cite{Frixione:2007nw,Nason:2004rx,Frixione:2007vw,Alioli:2010xd} serves as the benchmark signal MC sample and is normalized to the theoretical cross section of 7.4~pb for a top-quark mass of $m_{t}=172.5~\gevcc$~\cite{Czakon:2011xx}. We include hadronic $W$-boson decays of $\ttbar$ events, where one jet from bottom-quark hadronization or $W$ boson hadronic decay is misidentified as a charged lepton, in the background categories and estimate the contribution of this process with the \textsc{powheg} $\ttbar$ sample with the same normalization as the signal. The expected rates of background processes and the $\ttbar$ signal, together with the observed number of events in the signal region, are listed in Table~\ref{table:BackgroundEstimationSignal}. Excellent agreement is observed.

\begin{table}[htb]
\PRLBckEsti
\caption{Expected number of events in data corresponding to 9.1~$\fbm$ of integrated luminosity along with the observed number of events passing all event selections. The quoted uncertainties in each row are the quadratic sum of the statistical and systematic uncertainties, calculated in the same way as Ref.~\cite{Aaltonen:2013tsa}.}
\label{table:BackgroundEstimationSignal}
\end{table}

Assuming charge-parity symmetry in the strong interaction, the $\afblep$ is defined as
\begin{linenomath*}
\begin{equation}
	\afblep = \frac{N(\qeta > 0) - N(\qeta < 0)}{N(\qeta > 0) + N(\qeta < 0)}, 
	\label{eqn:afblep}
\end{equation} 
\end{linenomath*}
where $N$ is the number of leptons, $q_{\ell}$ is the lepton electric charge, and $\eta_{\ell}$ is its pseudorapidity. An NLO SM calculation with both quantum-chromodynamics effects and electroweak effects predicts $\afblep = 0.038\pm0.003$~\cite{Bernreuther:2012sx}. If the genuine value of $\afbtt$ would be that measured by the CDF collaboration~\cite{Aaltonen:2012it}, the predicted value for $\afblep$ for top quarks decaying according to the SM would be $0.070<\afblep<0.076$~\cite{Aaltonen:2013vaf}. Previous measurements of $\afblep$ in the lepton+jets final state by the CDF collaboration and in the lepton+jets and dilepton final state by the D0 collaboration yielded 0.094${}^{+0.032}_{-0.029}$~\cite{Aaltonen:2013vaf} and 0.047$\pm$0.027~\cite{Abazov:2013wxa,Abazov:2014oea}, respectively. A second observable, $\afbdeta$, can be defined in the dilepton final state analogously to $\afbtt$ as
\begin{linenomath*}
\begin{equation}
	\afbdeta = \frac{N(\Delta \eta > 0) - N(\Delta \eta < 0)}{N(\Delta \eta > 0) + N(\Delta \eta < 0)},
	\label{eqn:afbdeta}
\end{equation}
\end{linenomath*}
where $\Delta \eta = \eta_{\ell^+} - \eta_{\ell^-}$. The NLO SM prediction is $\afbdeta = 0.048 \pm 0.004$~\cite{Bernreuther:2012sx}. A measurement by the D0 collaboration in the dilepton final state is $\afbdeta = 0.123 \pm 0.056$~\cite{Abazov:2013wxa}.

We simulate $\ttbar$ production and decay in various plausible SM and beyond-SM scenarios to study the expected lepton pseudorapidity spectrum in a large range of $\afblep$ and $\afbdeta$ values. The \textsc{powheg} $\ttbar$ MC sample serves as a best estimate of the SM. It gives parton-level inclusive values of $\afblep = 0.024$ and $\afbdeta = 0.030$. These predictions are different from the NLO SM calculation in Ref.~\cite{Bernreuther:2012sx} since the simulation does not account for the electroweak corrections~\cite{kunhRodrigo,*Manohar2012313,*PhysRevD.84.093003}. Three $\ttbar$ MC samples that include a class of relatively light and wide axigluons (with masses at 200~$\gevcc$ and widths at $50~\gev$) with left-handed, right-handed, and axial axigluon couplings to the quarks~\cite{PhysRevD.87.034039} serve as benchmark simulation samples to model various SM extensions. Each predict an $\afbtt$ value similar to that observed by the CDF collaboration~\cite{Aaltonen:2012it}, but the polarization of the top quarks results in different values of $\afblep$ and $\afbdeta$. These samples are generated with \textsc{madgraph}~\cite{Alwall:2007st} and have $\afblep$ values of $-0.063$, $0.050$, and $0.151$ and $\afbdeta$ values of $-0.092$, $0.066$, and $0.218$, respectively.

Due to the limited detector coverage ($|\eta_{l}|<2.0$ for electrons and $|\eta_{l}|<1.1$ for muons), imperfect detector acceptance, and contamination from non-$\ttbar$ sources, a correction and extrapolation procedure is needed to determine the inclusive parton-level $\afblep$ from the data. Simulated samples show that the $\qeta$ distribution of the leptons at the parton level is modeled accurately by the sum of two Gaussian distributions with common means, and widths and proportions independent of the simulated model~\cite{DGPheno}. The asymmetry in each scenario arises from the shift of the mean of the $\qeta$ distribution. Using this knowledge, we follow a procedure that is similar to that described in Ref.~\cite{Aaltonen:2013vaf} to account for the detector coverage, detector acceptance and background effects described above. The $\qeta$ distribution of leptons is decomposed into a symmetric part and an asymmetric part as functions of $\qeta$ in the range $\qeta \geq 0$,
\begin{linenomath*}
\begin{subequations}
\begin{align}
	\mathcal{S}( \qeta ) &= \frac{\mathcal{N}(\qeta)+\mathcal{N}(-\qeta)}{2} \label{eqn:S}, \rm{and}\\
	\mathcal{A}( \qeta ) &= \frac{\mathcal{N}(\qeta)-\mathcal{N}(-\qeta)}{\mathcal{N}(\qeta)+\mathcal{N}(-\qeta)}\label{eqn:A},
\end{align}
\label{eqn:decomposition}
\end{subequations}
\end{linenomath*}
where $\mathcal{N}(\qeta)$ represents the number of events as a function of $\qeta$.  The differential contribution to the inclusive $\afblep$ as a function of $\qeta$ is calculated using the expression
\begin{linenomath*}
\begin{equation}
\frac{\mathcal{S}(\qeta) \times \mathcal{A}(\qeta)}{\int_{0}^{\infty}d(\qetaprime)~\mathcal{S}(\qetaprime)}\label{eqn:SA_Seq1},
\end{equation}
\end{linenomath*}
and the inclusive $\afblep$ defined in Eq.~(\ref{eqn:afblep}) is then written as the integral of Eq.~(\ref{eqn:SA_Seq1}),
\begin{linenomath*}
\begin{equation}
	\afblep 
	        = \frac{\int_{0}^{\infty}d(\qeta)~ [\mathcal{S}(\qeta) \times \mathcal{A}(\qeta)] }{\int_{0}^{\infty} d (\qetaprime) ~\mathcal{S}(\qetaprime)}. \label{eqn:S_A_AFB}
\end{equation}
\end{linenomath*}
The measurement methodology is simplified because the symmetric part of the $\qeta$ distributions at the parton level is very similar across models as the mean of the $\qeta$ distribution is always close to zero in all models and small compared to the width, which is always around unity. Hence, using the distribution from any simulated sample only introduces an uncertainty that is tiny compared to the dominant uncertainties. Additionally, the differential asymmetry described in Eq.~(\ref{eqn:A}) is readily measured and allows discrimination among models. We note that for $\qeta<2.5$, the differential asymmetry in Eq.~(\ref{eqn:A}) is modeled accurately by the simple functional form
\begin{linenomath*}
\begin{equation}
	\mathcal{A}( \qeta ) = a \cdot \mathrm{tanh}\left(\frac{\qeta}{2} \right)\label{eqn:tanh}
\end{equation}
\end{linenomath*}
where $a$ is a free parameter that is directly related to the asymmetry.

\begin{figure}[hbtp]
\begin{center}
        \includegraphics[width=\columnwidth]{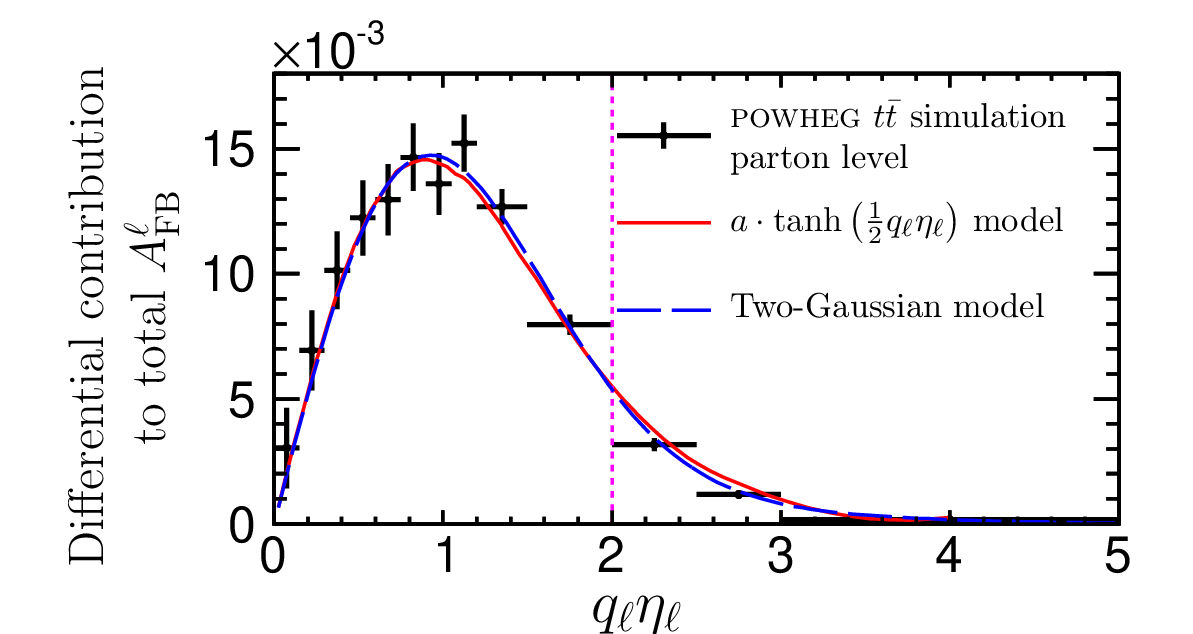}
        \caption{Differential contribution to $\afblep$ as a function of $\qeta$ for the \textsc{powheg} simulation of $\ttbar$ production. The solid curve shows the estimation with Eq.~(\ref{eqn:SA_Seq1}) where $\mathcal{A}(\qeta)$ is obtained with a fit of Eq.~(\ref{eqn:tanh}) on the asymmetric part of the $\qeta$ spectrum from the sample, and $\mathcal{S}(\qeta)$ is directly from the sample; the dashed curve is from the two-Gaussian model~\cite{DGPheno}. The vertical dashed line indicates the outer limits of the acceptance regions for charged leptons which is $|\qeta|=2.0$.}
        \label{fig:s_a_hepg_ttktfp}
\end{center}
\end{figure}

Figure~\ref{fig:s_a_hepg_ttktfp} shows the differential contribution to the inclusive $\afblep$ expected at parton level from the \textsc{powheg} simulation, along with comparisons with predictions from the two-Gaussian model and the simple functional form of Eq.~(\ref{eqn:tanh}). Both models describe the distribution accurately. The integral gives the total inclusive asymmetry, and the fraction of the unmeasured asymmetry where $|\qeta| > 2.0$ is approximately 11\%. The distributions for the various simulated samples, including the models listed above as well as those generated with \textsc{pythia}~\cite{Sjostrand:2006za} and \textsc{alpgen}~\cite{Mangano:2002ea}, show that the shapes are very similar, supporting the measurement methodology.

The strategy is to measure the shape of the asymmetric component of the data after background subtraction and use the symmetric component of the parton-level $\qeta$ distribution from the \textsc{powheg} $\ttbar$ sample to reproduce the inclusive parton-level value of $\afblep$. This method is checked for the wide variety of input $\afblep$ values using the fully simulated $\ttbar$ samples. For both the two-Gaussian model and the simplified functional form of Eq.~(\ref{eqn:tanh}), the method returns $\afblep$ values that are consistent with the parton-level inclusive values. We include an asymmetric-modeling systematic uncertainty of $\pm 0.006$, which covers any possible bias observed.

\begin{figure}[hbtp]
\begin{center}
	\subfigure[]{\includegraphics[width=\columnwidth]{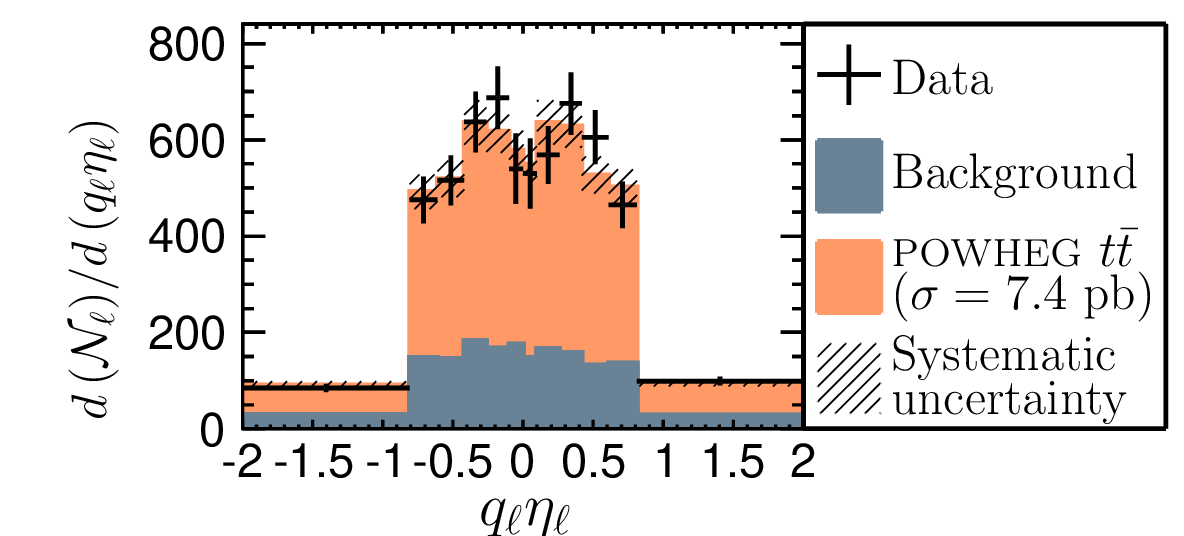}\label{fig:01_qeta_stack2}}\\
	\subfigure[]{\includegraphics[width=\columnwidth]{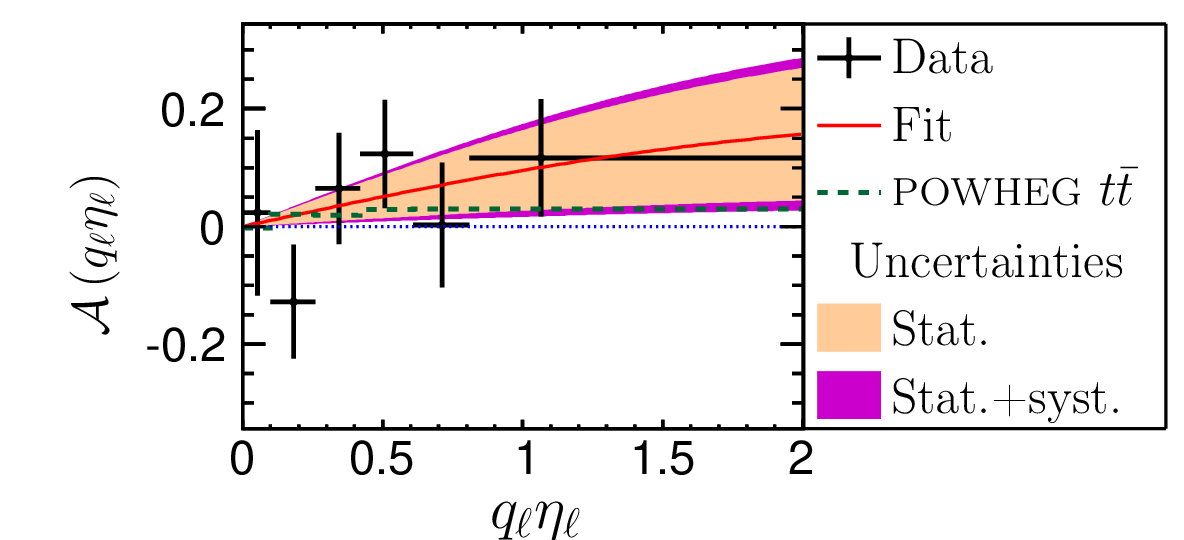}\label{fig:02_DATA_Fit}}
   \caption{(a): Comparison of the observed number of leptons as a function of $\qeta$ with the SM expectations. (b): Asymmetric part of the distribution in (a) defined in Eq.~(\ref{eqn:A}) from data after background subtraction together with the best fit with Eq.~(\ref{eqn:tanh}) and the expectations from the \textsc{powheg} MC model. The bands indicate the one standard deviation uncertainty for statistical and statistical+systematic uncertainties.}
   \label{fig:qetafigure}
\end{center}
\end{figure}

The observed number of events as a function of $\qeta$ is shown in Fig.~\ref{fig:qetafigure} along with the SM expectations from the $\ttbar$ signal and backgrounds. Figure~\ref{fig:02_DATA_Fit} shows the asymmetric component of the data after background subtraction along with the best fit description, which yields a value of $a = 0.21 \pm  0.15 (\rm{stat})$. Applying Eq.~(\ref{eqn:S_A_AFB}), we find $\afblep = 0.072 \pm 0.052 (\rm{stat})$.

The dominant source of systematic uncertainty is due to the background uncertainties and is estimated to be $\pm 0.029$ using pseudoexperiments, which covers both the uncertainties in the background normalizations and the uncertainties in modeling the $\afblep$ of the backgrounds. The next most important source of systematic uncertainty is the $\pm 0.006$ asymmetric-modeling contribution discussed above. The jet-energy-scale systematic uncertainty is estimated to be $\pm 0.004$ by varying the jet energies within their uncertainties. The variations obtained by using the symmetric model from various MC samples are assigned as the symmetric-modeling systematic uncertainty, which is $\pm 0.001$. Other sources of uncertainties due to the uncertainties in the parton showering model, the modeling of color reconnection, the amount of initial-state and final-state radiation, and the uncertainty on the parton distribution functions, are found to be negligible. The total systematic uncertainty, $\pm 0.03$, is estimated by summing the individual contributions in quadrature. The final result is $\afblep = 0.072 \pm 0.052 (\rm{stat}) \pm 0.030 (\rm{syst})$. This result is consistent with the NLO SM expectation, the measurement in the lepton+jets final state by the CDF collaboration~\cite{Aaltonen:2013vaf} and the combined measurement in both the lepton+jets and dilepton final state by the D0 collaboration~\cite{Abazov:2013wxa,Abazov:2014oea}.

\begin{figure}[hbtp]
\begin{center}
	\subfigure[]{\includegraphics[width=\columnwidth]{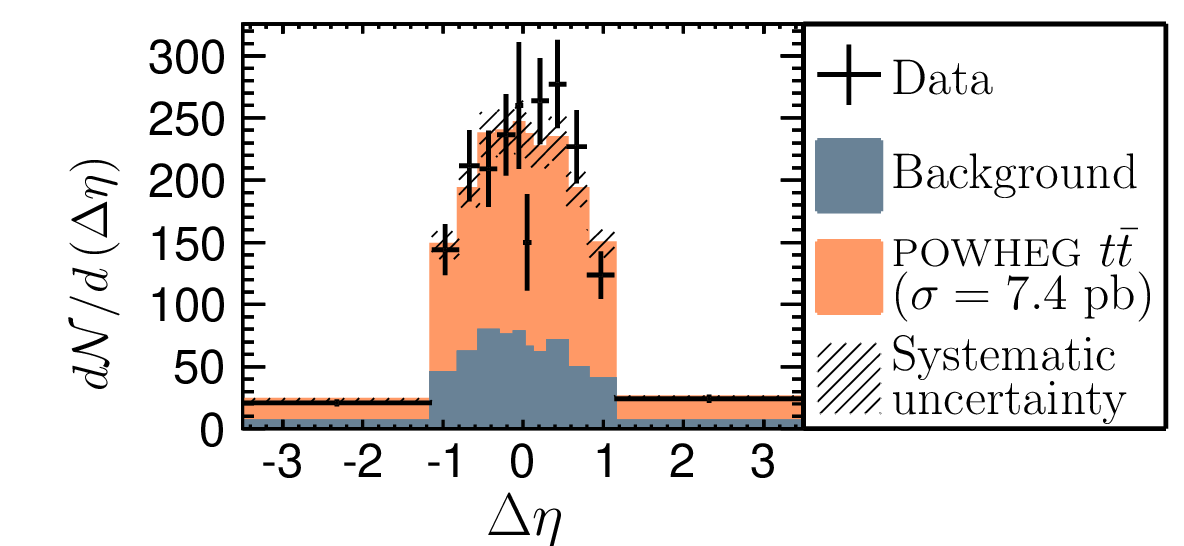}\label{fig:01_deta_stack2}}\\
	\subfigure[]{\includegraphics[width=\columnwidth]{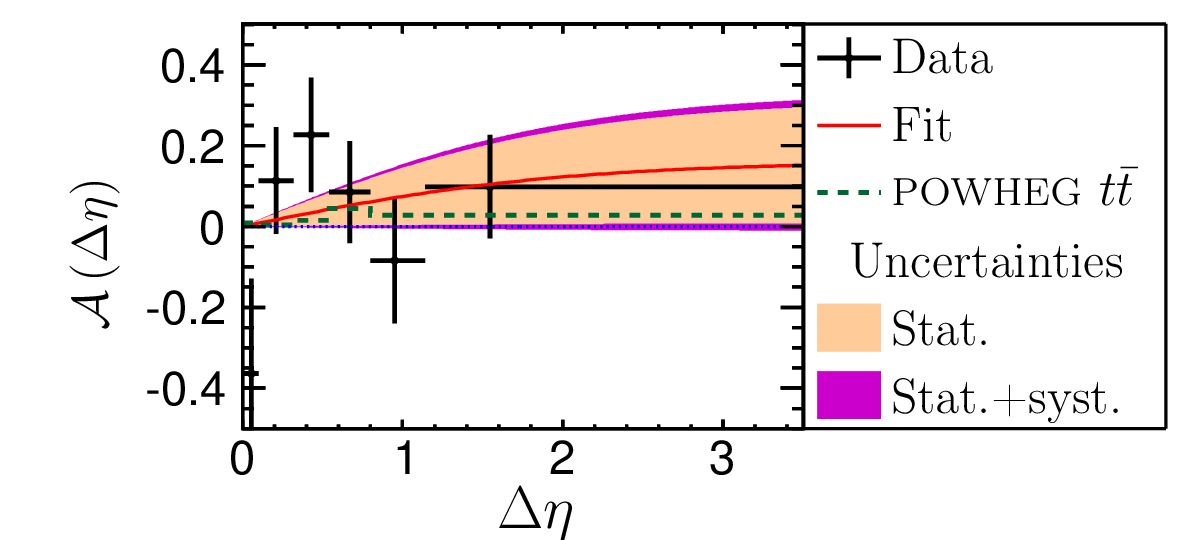}\label{fig:07_deta_DATA_Fit}}
   \caption{The same figures as Fig.~\ref{fig:qetafigure}, but as a function of $\deta$ instead of $\qeta$.}
   \label{fig:detafigure}
\end{center}
\end{figure}

Identical methodologies are used for measuring $\afbdeta$. The observed number of events as a function of $\deta$ is shown in Fig.~\ref{fig:detafigure}. We measure $a = 0.16 \pm  0.15 (\rm{stat})$ and $\afbdeta = 0.076 \pm 0.072 (\rm{stat}) \pm 0.039 (\rm{syst})$, where the dominant systematic uncertainty is from backgrounds and has a value of $\pm 0.037$. The asymmetric and symmetric-modeling systematic uncertainties are estimated to be $\pm 0.012$ and $\pm 0.004$, respectively. The jet-energy-scale systematic uncertainty is estimated to be $\pm 0.003$. Other systematic uncertainties are negligible. This result is consistent with both the NLO SM calculation~\cite{Bernreuther:2012sx} and the same measurement in the dilepton final state by the D0 collaboration~\cite{Abazov:2013wxa}.

In order to obtain a more sensitive measurement, we combine the dilepton measurement of $\afblep$ with the CDF measurement in the lepton+jets final state reported in Ref.~\cite{Aaltonen:2013vaf}, \(\afblep = 0.094 \pm 0.024 (\rm{stat}) ^{+0.022}_{-0.017} (\rm{syst}) \). The combination is based on the asymmetric iterative algorithm of the \textit{best linear unbiased estimates} approach~\cite{Lyons1988110,*PhysRevD.41.982,*Valassi2003391,Group:2008vx}. Since the measurements use statistically independent samples, the statistical uncertainties are uncorrelated. The background systematic uncertainties are treated as uncorrelated since they are mainly caused by the uncertainties in the modeling of the background $\qeta$ distributions, which are largely uncorrelated between the two measurements. The recoil-modeling systematic uncertainty in the lepton+jets measurement and the asymmetric-modeling systematic uncertainty in the dilepton measurement are both designed to cover the potential biases introduced by the measurement methodology, and are thus treated as fully correlated. The jet-energy-scale systematic uncertainties are also treated as fully correlated. The other systematic uncertainties are negligible in one of the two measurements, thus only the non-negligible part is included.

The combined result is $\afblep = 0.090^{+0.028}_{-0.026}$, where 80\% of the measurement weight is due to the lepton+jets result and 20\% is due to the dilepton result. The difference in the weights is mostly due to the larger size of the lepton+jets final state sample. The correlation between the two measurement uncertainties is estimated to be 2.6\%.

In conclusion, we measure the inclusive parton-level leptonic forward-backward asymmetry and leptonic pair asymmetry of top-quark pairs decaying into the dilepton final state using the full CDF Run II data set. The results are	$ \afblep = 0.072 \pm 0.060$ and $\afbdeta = 0.076 \pm 0.082$, both consistent with previous determinations and expectations. A combination of the CDF leptonic $\afb$ measurements yields $\afblep = 0.090^{+0.028}_{-0.026}$. This result is about two standard deviations larger than the NLO SM calculation of $\afblep = 0.038 \pm 0.003$~\cite{Bernreuther:2012sx}, but is consistent with the 0.070--0.076 range expected under the assumption of unpolarized top-quark production and SM top-quark decay, given the measured value of $\afbtt$ in the lepton+jets final state by the CDF collaboration~\cite{Aaltonen:2013vaf}. This result is also consistent with the $\afblep$ measured by the D0 collaboration~\cite{Abazov:2013wxa,Abazov:2014oea}.

\begin{acknowledgments}
We thank the Fermilab staff and the technical staffs of the
participating institutions for their vital contributions. This work
was supported by the U.S. Department of Energy and National Science
Foundation; the Italian Istituto Nazionale di Fisica Nucleare; the
Ministry of Education, Culture, Sports, Science and Technology of
Japan; the Natural Sciences and Engineering Research Council of
Canada; the National Science Council of the Republic of China; the
Swiss National Science Foundation; the A.P. Sloan Foundation; the
Bundesministerium f\"ur Bildung und Forschung, Germany; the Korean
World Class University Program, the National Research Foundation of
Korea; the Science and Technology Facilities Council and the Royal
Society, United Kingdom; the Russian Foundation for Basic Research;
the Ministerio de Ciencia e Innovaci\'{o}n, and Programa
Consolider-Ingenio 2010, Spain; the Slovak R\&D Agency; the Academy
of Finland; the Australian Research Council (ARC); and the EU community
Marie Curie Fellowship Contract No. 302103.
\end{acknowledgments}

\bibliographystyle{apsrev4-1-JHEPfix}
\bibliography{PubCitations}
\end{document}

%% file: afbcommands.tex
\newcommand{\MetVec}{\mbox{$\vec E\kern-0.50em\raise0.10ex\hbox{/}_{T}$}}
\newcommand{\MetVecRaw}{\mbox{$\vec E\kern-0.50em\raise0.10ex\hbox{/}^{raw}_{T}$}}
\newcommand{\Met}{\mbox{$E\kern-0.50em\raise0.10ex\hbox{/}_{T}$}}
\newcommand{\MetRaw}{\mbox{$E\kern-0.50em\raise0.10ex\hbox{/}_{T}^{raw}$}}
\newcommand{\MetSpec}{\mbox{$E\kern-0.50em\raise0.10ex\hbox{/}_{T}^{spec}$}}

\newcommand{\MetSig}{\mbox{$E\kern-0.50em\raise0.10ex\hbox{/}_{T}^{sig}$}}
\newcommand{\fbm}{\rm{fb}^{-1}}

\newcommand{\ttbar}{t \bar{t}}

\newcommand{\gev}{\rm{GeV}}
\newcommand{\gevc}{\textrm{GeV}/\textit{c}}
\newcommand{\gevcc}{\textrm{GeV}/\textit{c}^2}

\newcommand{\pt}{p_{T}}

\newcommand{\MET}{\mbox{$E\kern-0.50em\raise0.10ex\hbox{/}_{T}$}}
\newcommand{\met}{\mbox{$E\kern-0.50em\raise0.10ex\hbox{/}_{T}$}}
\newcommand{\METzero}{\mbox{$E\kern-0.50em\raise0.10ex\hbox{/}_{T}^0$}}
\newcommand{\vecMET}{\mbox{$\vec{E}\kern-0.50em\raise0.10ex\hbox{/}_{T}$}}
\newcommand{\METSPEC}{\mbox{$E\kern-0.50em\raise0.10ex\hbox{/}_{Tspec}$}}

\def\slantfrac#1#2{\kern.1em^{#1}\kern-.3em/\kern-.1em_{#2}}

\newcommand{\afb}{A_{\text{FB}}}

\newcommand{\qeta}{q_{\ell}\eta_{\ell}}
\newcommand{\qetaprime}{q_{\ell}'\eta_{\ell}'}
\newcommand{\deta}{\Delta\eta}
\newcommand{\afblep}{A_{\text{FB}}^{\ell}}
\newcommand{\afbdeta}{A_{\text{FB}}^{\ell\ell}}
\newcommand{\afbtt}{A_{\text{FB}}^{t\bar{t}}}

%% file: tables/PublicTables.tex
\newcommand{\PRLBckEsti}{
  \begin{tabular}{l r@{$\pm$}l}
\hline\hline
\multicolumn{1}{c}{ Source  }                        & \multicolumn{2}{c}{Events} \\
  \hline
Diboson                                         &   31&6 \\
$Z/\gamma^{*}$+jets                              &   50&6 \\
$W$+jets                                   &   64&17 \\
$\ttbar\text{ non-dilepton}$                    &   14.6&0.8 \\
  \hline
\multicolumn{1}{l}{ Total background}              &   160&21 \\
\multicolumn{1}{l}{$t\bar{t}$ ($\sigma = 7.4$ pb)} &   408&19 \\
  \hline
 Total SM expectation                                &   568&40 \\
  \hline
\multicolumn{1}{l}{ Observed  }                    &\multicolumn{2}{c}{569} \\
  \hline\hline
  \end{tabular}
}

%% file: author.tex
\affiliation{Institute of Physics, Academia Sinica, Taipei, Taiwan 11529, Republic of China}
\affiliation{Argonne National Laboratory, Argonne, Illinois 60439, USA}
\affiliation{University of Athens, 157 71 Athens, Greece}
\affiliation{Institut de Fisica d'Altes Energies, ICREA, Universitat Autonoma de Barcelona, E-08193, Bellaterra (Barcelona), Spain}
\affiliation{Baylor University, Waco, Texas 76798, USA}
\affiliation{Istituto Nazionale di Fisica Nucleare Bologna, \ensuremath{^{ii}}University of Bologna, I-40127 Bologna, Italy}
\affiliation{University of California, Davis, Davis, California 95616, USA}
\affiliation{University of California, Los Angeles, Los Angeles, California 90024, USA}
\affiliation{Instituto de Fisica de Cantabria, CSIC-University of Cantabria, 39005 Santander, Spain}
\affiliation{Carnegie Mellon University, Pittsburgh, Pennsylvania 15213, USA}
\affiliation{Enrico Fermi Institute, University of Chicago, Chicago, Illinois 60637, USA}
\affiliation{Comenius University, 842 48 Bratislava, Slovakia; Institute of Experimental Physics, 040 01 Kosice, Slovakia}
\affiliation{Joint Institute for Nuclear Research, RU-141980 Dubna, Russia}
\affiliation{Duke University, Durham, North Carolina 27708, USA}
\affiliation{Fermi National Accelerator Laboratory, Batavia, Illinois 60510, USA}
\affiliation{University of Florida, Gainesville, Florida 32611, USA}
\affiliation{Laboratori Nazionali di Frascati, Istituto Nazionale di Fisica Nucleare, I-00044 Frascati, Italy}
\affiliation{University of Geneva, CH-1211 Geneva 4, Switzerland}
\affiliation{Glasgow University, Glasgow G12 8QQ, United Kingdom}
\affiliation{Harvard University, Cambridge, Massachusetts 02138, USA}
\affiliation{Division of High Energy Physics, Department of Physics, University of Helsinki, FIN-00014, Helsinki, Finland; Helsinki Institute of Physics, FIN-00014, Helsinki, Finland}
\affiliation{University of Illinois, Urbana, Illinois 61801, USA}
\affiliation{The Johns Hopkins University, Baltimore, Maryland 21218, USA}
\affiliation{Institut f\"{u}r Experimentelle Kernphysik, Karlsruhe Institute of Technology, D-76131 Karlsruhe, Germany}
\affiliation{Center for High Energy Physics: Kyungpook National University, Daegu 702-701, Korea; Seoul National University, Seoul 151-742, Korea; Sungkyunkwan University, Suwon 440-746, Korea; Korea Institute of Science and Technology Information, Daejeon 305-806, Korea; Chonnam National University, Gwangju 500-757, Korea; Chonbuk National University, Jeonju 561-756, Korea; Ewha Womans University, Seoul, 120-750, Korea}
\affiliation{Ernest Orlando Lawrence Berkeley National Laboratory, Berkeley, California 94720, USA}
\affiliation{University of Liverpool, Liverpool L69 7ZE, United Kingdom}
\affiliation{University College London, London WC1E 6BT, United Kingdom}
\affiliation{Centro de Investigaciones Energeticas Medioambientales y Tecnologicas, E-28040 Madrid, Spain}
\affiliation{Massachusetts Institute of Technology, Cambridge, Massachusetts 02139, USA}
\affiliation{University of Michigan, Ann Arbor, Michigan 48109, USA}
\affiliation{Michigan State University, East Lansing, Michigan 48824, USA}
\affiliation{Institution for Theoretical and Experimental Physics, ITEP, Moscow 117259, Russia}
\affiliation{University of New Mexico, Albuquerque, New Mexico 87131, USA}
\affiliation{The Ohio State University, Columbus, Ohio 43210, USA}
\affiliation{Okayama University, Okayama 700-8530, Japan}
\affiliation{Osaka City University, Osaka 558-8585, Japan}
\affiliation{University of Oxford, Oxford OX1 3RH, United Kingdom}
\affiliation{Istituto Nazionale di Fisica Nucleare, Sezione di Padova, \ensuremath{^{jj}}University of Padova, I-35131 Padova, Italy}
\affiliation{University of Pennsylvania, Philadelphia, Pennsylvania 19104, USA}
\affiliation{Istituto Nazionale di Fisica Nucleare Pisa, \ensuremath{^{kk}}University of Pisa, \ensuremath{^{ll}}University of Siena, \ensuremath{^{mm}}Scuola Normale Superiore, I-56127 Pisa, Italy, \ensuremath{^{nn}}INFN Pavia, I-27100 Pavia, Italy, \ensuremath{^{oo}}University of Pavia, I-27100 Pavia, Italy}
\affiliation{University of Pittsburgh, Pittsburgh, Pennsylvania 15260, USA}
\affiliation{Purdue University, West Lafayette, Indiana 47907, USA}
\affiliation{University of Rochester, Rochester, New York 14627, USA}
\affiliation{The Rockefeller University, New York, New York 10065, USA}
\affiliation{Istituto Nazionale di Fisica Nucleare, Sezione di Roma 1, \ensuremath{^{pp}}Sapienza Universit\`{a} di Roma, I-00185 Roma, Italy}
\affiliation{Mitchell Institute for Fundamental Physics and Astronomy, Texas A\&M University, College Station, Texas 77843, USA}
\affiliation{Istituto Nazionale di Fisica Nucleare Trieste, \ensuremath{^{qq}}Gruppo Collegato di Udine, \ensuremath{^{rr}}University of Udine, I-33100 Udine, Italy, \ensuremath{^{ss}}University of Trieste, I-34127 Trieste, Italy}
\affiliation{University of Tsukuba, Tsukuba, Ibaraki 305, Japan}
\affiliation{Tufts University, Medford, Massachusetts 02155, USA}
\affiliation{University of Virginia, Charlottesville, Virginia 22906, USA}
\affiliation{Waseda University, Tokyo 169, Japan}
\affiliation{Wayne State University, Detroit, Michigan 48201, USA}
\affiliation{University of Wisconsin, Madison, Wisconsin 53706, USA}
\affiliation{Yale University, New Haven, Connecticut 06520, USA}

\author{T.~Aaltonen}
\affiliation{Division of High Energy Physics, Department of Physics, University of Helsinki, FIN-00014, Helsinki, Finland; Helsinki Institute of Physics, FIN-00014, Helsinki, Finland}
\author{S.~Amerio\ensuremath{^{jj}}}
\affiliation{Istituto Nazionale di Fisica Nucleare, Sezione di Padova, \ensuremath{^{jj}}University of Padova, I-35131 Padova, Italy}
\author{D.~Amidei}
\affiliation{University of Michigan, Ann Arbor, Michigan 48109, USA}
\author{A.~Anastassov\ensuremath{^{v}}}
\affiliation{Fermi National Accelerator Laboratory, Batavia, Illinois 60510, USA}
\author{A.~Annovi}
\affiliation{Laboratori Nazionali di Frascati, Istituto Nazionale di Fisica Nucleare, I-00044 Frascati, Italy}
\author{J.~Antos}
\affiliation{Comenius University, 842 48 Bratislava, Slovakia; Institute of Experimental Physics, 040 01 Kosice, Slovakia}
\author{G.~Apollinari}
\affiliation{Fermi National Accelerator Laboratory, Batavia, Illinois 60510, USA}
\author{J.A.~Appel}
\affiliation{Fermi National Accelerator Laboratory, Batavia, Illinois 60510, USA}
\author{T.~Arisawa}
\affiliation{Waseda University, Tokyo 169, Japan}
\author{A.~Artikov}
\affiliation{Joint Institute for Nuclear Research, RU-141980 Dubna, Russia}
% ...

\collaboration{CDF Collaboration}
\altaffiliation[With visitors from]{
\ensuremath{^{a}}University of British Columbia, Vancouver, BC V6T 1Z1, Canada,
\ensuremath{^{b}}Istituto Nazionale di Fisica Nucleare, Sezione di Cagliari, 09042 Monserrato (Cagliari), Italy,
\ensuremath{^{c}}University of California Irvine, Irvine, CA 92697, USA,
\ensuremath{^{d}}Institute of Physics, Academy of Sciences of the Czech Republic, 182~21, Czech Republic,
\ensuremath{^{e}}CERN, CH-1211 Geneva, Switzerland,
\ensuremath{^{f}}Cornell University, Ithaca, NY 14853, USA,
\ensuremath{^{g}}University of Cyprus, Nicosia CY-1678, Cyprus,
\ensuremath{^{h}}Office of Science, U.S. Department of Energy, Washington, DC 20585, USA,
\ensuremath{^{i}}University College Dublin, Dublin 4, Ireland,
\ensuremath{^{j}}ETH, 8092 Z\"{u}rich, Switzerland,
\ensuremath{^{k}}University of Fukui, Fukui City, Fukui Prefecture, Japan 910-0017,
\ensuremath{^{l}}Universidad Iberoamericana, Lomas de Santa Fe, M\'{e}xico, C.P. 01219, Distrito Federal,
\ensuremath{^{m}}University of Iowa, Iowa City, IA 52242, USA,
\ensuremath{^{n}}Kinki University, Higashi-Osaka City, Japan 577-8502,
\ensuremath{^{o}}Kansas State University, Manhattan, KS 66506, USA,
\ensuremath{^{p}}Brookhaven National Laboratory, Upton, NY 11973, USA,
\ensuremath{^{q}}Queen Mary, University of London, London, E1 4NS, United Kingdom,
\ensuremath{^{r}}University of Melbourne, Victoria 3010, Australia,
\ensuremath{^{s}}Muons, Inc., Batavia, IL 60510, USA,
\ensuremath{^{t}}Nagasaki Institute of Applied Science, Nagasaki 851-0193, Japan,
\ensuremath{^{u}}National Research Nuclear University, Moscow 115409, Russia,
\ensuremath{^{v}}Northwestern University, Evanston, IL 60208, USA,
\ensuremath{^{w}}University of Notre Dame, Notre Dame, IN 46556, USA,
\ensuremath{^{x}}Universidad de Oviedo, E-33007 Oviedo, Spain,
\ensuremath{^{y}}CNRS-IN2P3, Paris, F-75205 France,
\ensuremath{^{z}}Universidad Tecnica Federico Santa Maria, 110v Valparaiso, Chile,
\ensuremath{^{aa}}The University of Jordan, Amman 11942, Jordan,
\ensuremath{^{bb}}Universite catholique de Louvain, 1348 Louvain-La-Neuve, Belgium,
\ensuremath{^{cc}}University of Z\"{u}rich, 8006 Z\"{u}rich, Switzerland,
\ensuremath{^{dd}}Massachusetts General Hospital, Boston, MA 02114 USA,
\ensuremath{^{ee}}Harvard Medical School, Boston, MA 02114 USA,
\ensuremath{^{ff}}Hampton University, Hampton, VA 23668, USA,
\ensuremath{^{gg}}Los Alamos National Laboratory, Los Alamos, NM 87544, USA,
\ensuremath{^{hh}}Universit\`{a} degli Studi di Napoli Federico I, I-80138 Napoli, Italy
}
\noaffiliation

%% file: TopDileptonAfb_PRL.bbl
\begin{thebibliography}{52}
\bibitem [{geo()}]{geometry}%
  \BibitemOpen
  \href@noop {} {}\bibinfo {note} {We use a cylindrical coordinate system with
  the origin at the center of the CDF II detector, $z$ pointing in the
  direction of the proton beam, $\theta$ and $\phi$ representing the polar and
  azimuthal angles, respectively, and pseudorapidity defined by $\eta = - \rm
  ln$ $\rm tan(\theta/2)$. The transverse momentum $p_T$ (transverse energy
  $E_{T}$) is defined to be $p\sin\theta$ ($E\sin\theta$).}\BibitemShut {Stop}%
\bibitem [{MET()}]{MET}%
  \BibitemOpen
  \href@noop {} {}\bibinfo {note} {The missing transverse energy $\met$ is
  defined to be $-\Sigma_{i}E_{T}^{i}$\^{n}$_{i}$ where $i$ is the calorimeter
  tower number with $|\eta|<3.6$, \^{n}$_{i}$ is a unit vector perpendicular to
  the beam axis and pointing at the $i$th calorimeter tower.}\BibitemShut
  {Stop}%
\end{thebibliography}
